**Space Weather: From Solar Origins to Risks and Hazards Evolving in Time**

**Natalia Buzulukova[1,2]\*, Bruce Tsurutani[3]**

[1]NASA GSFC, Heliophysics Division, Geospace Physics Laboratory, Greenbelt, MD, USA

[2]University of Maryland, Department of Astronomy, College Park, MD, USA

[3]Heliospheric Physics and Astrophysics Section, Jet Propulsion Laboratory, California Institute of Technology, Pasadena, CA, USA

**\* Correspondence:**
Natalia Buzulukova
nbuzulukova@gmail.com

**Review paper**

**Abstract**

Space Weather is the portion of space physics that has a direct effect on humankind. Space Weather is an old branch of space physics that originates back to 1808 with the publication of a paper by the great naturalist Alexander von Humboldt (von Humboldt, 1808), first defining a "Magnetische Ungewitter" or magnetic storm from auroral observations from his home in Berlin, Germany. Space Weather is currently experiencing explosive growth, because its effects on human technologies have become more and more diverse. Space Weather is due to the variability of solar processes that cause interplanetary, magnetospheric, ionospheric, atmospheric and ground level effects. Space Weather can at times have strong impacts on technological systems and human health. The threats and risks are not hypothetical, and in the event of extreme Space Weather events the consequences could be quite severe for humankind. The purpose of the review is to give a brief overall view of the full chain of physical processes responsible for Space Weather risks and hazards, tracing them from solar origins to effects and impacts in interplanetary space, in the Earth's magnetosphere and ionosphere and at the ground. In addition, the paper shows that the risks associated with Space Weather have not been constant over time; they have evolved as our society becomes more and more technologically advanced. The paper begins with a brief introduction to the Carrington event, arguably the greatest geomagnetic storm in recorded history. Next, the descriptions of the strongest known Space Weather processes are reviewed, tracing them from their solar origins. The concepts of geomagnetic storms and substorms are briefly introduced. The main effects/impacts of Space Weather are also considered, including geomagnetically induced currents (GICs) which are thought to cause power outages. The effects of radiation on avionics and human health, ionospheric effects and impacts, and thermosphere effects and satellite drag will also be discussed. Finally, we will discuss the current challenges of Space Weather forecasting and examine some of the worst-case scenarios.

**Type: Review Paper**

# Space Weather: from solar origins to risks and hazards evolving in time

## Introduction

The fields of Space Weather and space physics were strongly stimulated by the publication of a paper by the great naturalist Alexander von Humboldt (1808). Von Humboldt observed that oscillations of magnetic needles occurred for over 4 hrs when there were auroras overhead at his home in Berlin, Germany. He called this a "Magnetische Ungewitter" or "magnetic storm". Beginning in the mid-19th century, many technological systems on Earth were impacted by Space Weather. Movements of the telegraph magnetic needles of the Midland Railroad (England) were observed to be coincident with auroral sightings (Barlow, 1849).  On 1 September 1859, R. C. Carrington and R. Hodgson both observed and reported an optical solar flare (Carrington, 1859; Hodgson, 1859). Carrington also noted that a large magnetic storm on Earth occurred some 17 hours and 40 minutes after the flare. During the magnetic storm, fires were started in telegraph stations (Loomis, 1861; see also Shea and Smart, 2006 for a compilation of 8 works by Loomis). Details of this extreme magnetic storm have been reported by Kimball (1960), Tsurutani et al. (2003) and Lakhina et al. (2012). The special issue of Advances in Space Research (Clauer and Siscoe, 2006) brings together 23 articles on the Carrington event**.**

People were aware about geomagnetic activity long before the Carrington event (Maden, 2020). Although sailors of the past did not know what geomagnetic storms or substorms were, geomagnetic activity was of great interest to nations with sailing ships. Geomagnetic storms are worldwide events that can last from hours to tens of hours (Chapman and Bartels, 1940; Gonzalez et al., 1994). The Earth's external magnetic fields (the magnetosphere) becomes filled with energetic particles. The circulation of these particles, called the "ring current" can change the Earth's surface magnetic field directions by up to ~5˚. S.-I. Akasofu, as a student of S. Chapman, described a sequence of spatial auroral evolution from all-sky camera data (Akasofu, 1964)[1]. Because he believed this sequence to be a fundamental sub-element of a magnetic storm**,** Akasofu called this a "substorm".

A substorm is a sequence of auroral features lasting from ~30 min to an hour. Not only are intense auroras associated with substorms, but strong ionospheric currents are as well. The relationship between substorms and magnetic storms still remains a major topic of study today. After decades of intense research, there is still no definite answer.

More recently in the space age satellite operators and designers have had to address effects of Space Weather on a daily basis. Power grid operators closely monitor geomagnetically induced currents (GICs) that could damage/destroy transformers and grid infrastructure. Space Weather has many adverse effects on human health. Future crewed missions to Mars and the Moon view particle radiation storms as potentially life-threatening events. If an extreme Space Weather event occurs, the consequences for Earth could be truly problematic for humankind. Cannon et al. (2013) and Hapgood et al. (2021) estimated that the time it would take to repair damage to the UK power grid system from an extreme geomagnetic storm would be weeks to months. Oughton et al. (2017) have estimated that the worldwide losses due to an extreme magnetic storm would be between $7B and $42B per day.

---

[1] Kristian Birkeland published a two volume book entitled "The Norwegian Aurora Polaris Expedition 1902-1903" where he introduced "Polar Elementary Storms" (Birkeland, 1908). Birkeland used his terrella experiments to show that "corpuscular rays" from the Sun could penetrate into the polar regions of the Earth. Analysis of Birkeland's contribution can be found in Egeland and Burke (2010). Akasofu's 1964 work was the first to introduce substorm morphology and provide a modern definition of substorm phases.

**Space Weather: from solar origins to risks and hazards evolving in time**

Because of the diverse and complex nature of Space Weather effects, it is impossible to provide detailed descriptions of all the (presently known) primary facets of Space Weather here. The purpose of the review is to briefly give an overview of the full chain of physical processes responsible for Space Weather risks and hazards. We will trace them from solar origins to particular effects and impacts in interplanetary space and at Earth. Further reading is encouraged to understand the multiple and sometimes interrelated nature of Space Weather effects/impacts. Helpful reviews and in-depth analyses can be found in the following books: "Collisionless Shocks in the Heliosphere: A Tutorial Review" by Stone and Tsurutani (1985), "Collisionless Shocks in the Heliosphere: Reviews of Current Research" by Tsurutani and Stone (1985), "From the Sun: Auroras, Magnetic Storms, Solar Flares, Cosmic Rays" by Suess and Tsurutani (1998), "Magnetic Storms" by Tsurutani et al. (1997a), "Storm-Substorm Relationship" by Sharma et al. (2004), "Space Weather: The Physics Behind a Slogan" by Scherer et al. (2005), "Recurrent Magnetic Storms: Corotating Solar Wind Streams" by Tsurutani et al. (2006a), "The Sun and Space Weather" by Hanslmeier (2007), "Space Weather - Physics and Effects" by Bothmer and Daglis (2007), "Physics of Space Storms: From the Solar Surface to the Earth" by Koskinen (2011), "Extreme Events in Geospace: Origins, Predictability and Consequences" by Buzulukova (2018), "Space Weather Effects and Applications" by Coster et al.(2021), and a fairly comprehensive review of space plasma physics by Tsurutani et al. (2022a). We encourage the reader to explore the literature for the details of all physical processes.

## 1. Carrington event as a first encounter of the widespread effect of space weather on human technologies

The best-known Space Weather event is a chain of physical processes in August-September 1859, which began with the appearance on the Sun's surface of a group of unusually large sunspots (Carrington, 1859). The event culminated in arguably the largest geomagnetic storm in recorded history on September 2[nd] (Chapman and Bartels, 1940; Kimball, 1960; Tsurutani et al., 2003; Lakhina et al., 2005, 2012). This is now known as the 'Carrington event" or "Carrington storm.'

The appearance of bright and vibrant auroral displays such as in Figure 1a across much of the globe during the Carrington magnetic storm must have been a memorable if not terrifying experience. for many people (Kimball, 1960). There were dozens of eyewitness accounts describing many auroral forms that lasted for days. We now understand that magnetospheric and auroral ionospheric currents are parts of the giant current system, the so called Global Electric Circuit, that connects the magnetosphere ionosphere/upper atmosphere and the ground. The charged particles that carry out these currents (electrons and protons, with electrons being a primarily source of visible auroral forms) are accelerated at the distances 4,000-12,000 km above the Earth's surface and collide with neutral atoms in the topside atmosphere. These collisions mainly take place at 80-600 km altitude depending on the energy of the incident interacting charged particles (Shelley, 1995). As a result of electron collisional excitation of oxygen and nitrogen atoms and molecules, visible light is emitted with green, red, purple and blue colors, giving an observer a spectacular view of the dancing light of the Aurora Borealis in the Northern hemisphere, and the Aurora Australis in the Southern hemisphere (Figure 1b). A good review of auroral processes can be found in Paschmann et al. (2003).

Auroras also emit radiation in the ultraviolet and X-ray parts of the spectrum, but because these emissions are absorbed efficiently by the Earth's atmosphere, ultraviolet and X-ray auroras could only be observed from space or rocket experiments (Meier, 1991; Torr et al., 1995; Petrinec et al., 1999; Frey et al., 2004) and high-altitude balloon experiments (Winckler et al. 1959; Anderson and Milton, 1962; Brown et al. 1965; Millan et al. 2011), respectively.

# Space Weather: from solar origins to risks and hazards evolving in time

Evidence of unusually high auroral activity is scattered through mythology and ancient documents, but it is hard to decode the details from these old writings. At the time people did not understand the aurora and its causes (Mader, 2020). It was also difficult to understand that auroral lights might have significant impacts on human society in the future, other than psychological effects. Examples could be found in the essay "Auroral Omens of the American Civil War" (Love, 2014).

However, it is not just great auroral displays that draw the attention of scientists and the general public to the Carrington event. There were numerous documented anomalies related primarily to telegraph communications, the top technological system of the day. These included large voltages at the ends of telegraph wires and fires were started by arcing. More importantly, there were multiple interruptions in communication in Europe and North America, lasting from a few hours to most of the day on September 2nd, 1859. These disturbances were correctly attributed to the effect of "auroral electricity flows", although at that time scientists were unaware that strong ionospheric currents flowed parallel to Earth's surface at auroral latitudes at approximately 100 km altitude. These 'invisible' ionospheric currents are today called the "auroral electrojets". These currents can be as intense as one million Amperes. The precipitating auroral particles provide high conductivity by producing ionospheric ionization for the electrojets and the currents close the giant current loops/circuits connecting the ionosphere and the magnetosphere.

A detailed description of the telegraph effects such as fires set in telegraph offices in August/September 1859 can be found in the review by Boteler (2006). Thus, one could describe the Carrington event as a *Space Weather* event, where a blob of plasma and embedded magnetic fields called a Coronal Mass Ejection (CME) came from the Sun and impacted the Earth's magnetosphere (CMEs will be described in more detail later) and caused a series of adverse effects on human technologies: in this case and time, on the global system of telegraph communications. In modern terminology, the physical effects that were responsible for communication interruptions are called Geomagnetically Induced Currents (GICs). GICs are widely recognized today as one of the greatest dangers associated with Space Weather (see Section 4.1 for the details). As some observers of the Carrington event correctly recognized, the changes in auroral emission intensities were related to variations in the magnetic field direction and strength at the Earth's surface. The ionospheric auroral electrojets resulted in strong induced currents in Earth-bound conducting communication devices, such as telegraph wires. It is the strength of the effects and the widespread interruptions of services that make the Carrington event a special one, thus defining it as an example of Extreme Space Weather. A review by Lanzerotti (2017) presented other examples of Extreme Space Weather related to the occurrence of GICs, including the May 15 of 1921, March 24 of 1940, February 10 of 1958 magnetic storms and other events.

In 1859, humanity encountered the negative manifestations of Space Weather effects with the high technology at the time: telegraph communications. With the invention of the first radio devices in 1895 (independently by the Italian scientist G. Marconi and Russian scientist A. Popov), the demonstration of the first transatlantic transmission of radio waves in 1901 by Marconi and his team brought humanity into a new era of wireless communications. Since the propagation of radio waves over long distances depends on ionospheric conditions, this new technology became sensitive to new classes of Space Weather effects (see Section 4.4 of this Chapter for details).

The launch of the first Sputnik in 1957 marked the start of the Space Age. Today, the functioning of many government services and private sector enterprises, depend heavily on modern satellite telecommunication systems. A well-known example is the Global Navigation Satellite System (GNSS) which uses satellites to provide positioning, navigation, and timing (PNT) services with high accuracy. An incomplete list of GNSS applications includes airplane, ship and land navigation systems, mapping, surveying, emergency location-based services, and high-speed financial trading systems. In addition to government owned and operated GNSS systems, there is a diverse private sector of telecommunications companies around the world specialized in providing worldwide voice and data communication.

As satellite telecommunication services are closely integrated into all aspects of the life of modern society, it is necessary to understand the effects and negative impact of Space Weather events on the functionality and reliability of these systems. Factors that can affect the performance of a single satellite or deteriorate the whole system strongly depend on the satellite orbits and in general include:

- Particle radiation and plasma effects on satellites (Section 4.2);
- Ionospheric Space Weather effects (Section 4.4);
- Thermosphere (upper atmosphere) effects and satellite drag (Section 4.5).

How many satellites are there in Earth orbit at the moment? The US Government Accountability Office (GAO) website (https://www.gao.gov/products/gao-22-105166) estimates the number of operational satellites orbiting Earth at nearly 5,500 in the spring of 2022. The UCS database (https://www.ucsusa.org/resources/satellite-database) indicates that the number of currently operational satellites is 5,465 as of May 1, 2022. It appears that most of these satellites are privately owned. As of May 28, 2022, the American company SpaceX has launched 2,653 Starlink satellites (https://spaceflightnow.com/2022/05/18/falcon-9-starlink-4-18-live-coverage/)**.** In 2018, the United States Federal Communication Commission (FCC) authorized SpaceX to launch 4,425 satellites in Low Earth Orbit (LEO). As stated in the FCC press release, this authorization and similar authorizations are granted to a number of companies "… to provide broadband services using satellite technology that holds promise to expand Internet access, particularly in remote and rural areas across the country." (https://www.fcc.gov/document/fcc-authorizes-spacex-provide-broadband-satellite-services). It is worth noting that Starlink Low Earth Orbit (LEO) trajectories cover the critical altitudes of 500-600 km where satellite drag effects are important (to be discussed in Section 4.5). A recent example is SpaceX's loss of approximately 40 Starlink spacecraft (s/c) launched on February 3, 2022 and hit by two geomagnetic storms on February 3-4, 2022 (see SpaceX "Updates" at https://www.spacex.com/ updates / from February 8, 2022; Dang et al. 2022; Fang et al. 2022 and Tsurutani et al. 2022b). With the daily increase in demand for high-speed data communications, the number of satellites will likely increase substantially each year. Society's exposure to the risks associated with Space Weather will obviously increase as well.

## 2. Solar Origins of Space Weather

Space Weather is now a growing multi-disciplinary field that studies the origins of the variability of solar-terrestrial processes and links these processes with particular impacts on technological systems and human health. Because of Space Weather's diversity, it is important to understand the physical processes responsible for the most significant effects and impacts of Space Weather. Thus we need to understand space plasma physics even better because of Space Weather effects.

Figure 2 highlights the relationships between diverse Space Weather processes originating from the Sun and Space Weather impacts. Some of these processes are created and evolve in interplanetary space. The two predominant solar sources of Space Weather are active regions (ARs, see review by van Driel-Gesztelyi et al., 2015), decayed ARs and coronal holes (CHs, see review by Cranmer, 2009). ARs are a complex set of varying polarity sunspot magnetic fields in the photosphere of the Sun. The varying magnetic polarities are interconnected with the magnetic loops protruding above the surface. Due to evolution of the AR substructure and underlying subsurface dynamics, changes in the magnetic field interconnections (magnetic reconnection) lead to magnetic annihilation and sudden energy release in the form of a "solar flare" (Shibata and Magara, 2011; Toriumi and Wang, 2019). The solar flare photons cover the whole spectral scale from X-rays to UV to visible light photons (X-rays have to be detected by detectors onboard satellites or rockets above the Earth's atmosphere - thus only known during the space age). The visible light from a solar flare noted by Carrington (1859) and Hodgson (1859) is reasonably rare and thus as far as we know, was not reported in ancient documents. The magnetic reconnection at the Sun could also release a coronal mass ejection (CME), which will be described later. ARs occur primarily during the ~11-year solar cycle maximum sunspot number phase, called "solar maximum" for short (for details see the review by Hathaway, 2015 on solar cycle).

CHs are "dark regions" in the Sun seen in soft X-rays (Zilker, 1977; Suess, 1979; Harley and Sheeley, 1979). Therefore, before the space age there was no visible feature on the Sun that could explain a possible 27 day periodicity in geomagnetic activity at the Earth (Maunder, 1905, 1906) connecting the geomagnetic activity to the solar rotation[2]. However, Chree (1913) showed mathematically that the apparent periodicity was statistically significant and there was something on the Sun that was causing this.CHs are regions that have open magnetic fields (see review by Cranmer, 2009). It is known from Ulysses observations (Phillips et al. 1995) that the fast solar wind with speeds of ~750 to 800 km/s emanate from CHs. It is also known that Alfvén wave magnetic field fluctuations in the high speed streams (HSSs) cause substorm geomagnetic activity at Earth (Tsurutani et al. 2006b).

The solar origins of most Space Weather effects are mainly related to a few broad classes of phenomena: i) Solar eruptive phenomena (CMEs, solar flares and erupt filaments) which are linked to the dynamic processes of magnetic field reconfiguration (magnetic reconnection) and energy release in ARs and decayed ARs; (ii) variability of solar irradiance due to magnetic convection and 27-day solar rotation, including important spectral bands of extreme ultraviolet (EUV, 10-120 nm), and ultraviolet (UV, 120-400 nm) emissions; iii) solar coronal holes from which high speed streams (HSSs) of solar wind emanate. The HSSs interact with slow speed streams in interplanetary space forming large scale compressive structures called Corotating Interaction Regions (CIRs) (Smith and Wolf, 1976).

CIRs are responsible for CIR-driven types of geomagnetic storms. These are lower intensity storms than CME storms. The HSSs of ~750 to 800 km/s and embedded Alfvén waves cause auroral substorms and enhancement of magnetospheric electron fluxes with MeV energies (see Section 3.2 for details). CIRs and HSSs tend to occur during the declining phase of the solar cycle (Tsurutani et al., 2006b; Hajra et al. 2013).

---

[2] Although Edward E. Maunder is listed as the sole author of these two works, there is growing appreciation that his wife, astronomer and mathematician Annie S.D. Maunder, also contributed to the discovery of the statistical association between recurrent magnetic storms and solar activity. Please see for more details: https://mathshistory.st-andrews.ac.uk/Biographies/Maunder/

**Space Weather: from solar origins to risks and hazards evolving in time**

CMEs are giant magnetized plasma structures expelled from the Sun (Figure 3). CMEs could eject billions of tons of the Sun's material, and travel at speeds ranging from a few hundreds to a few thousands of km/sec (Forbes, 2000; Chen, 2011). The size of a single CME expands as it propagates from the Sun through the interplanetary medium and could reach ~0.5 astronomical unit (au) in width or larger by the time it reaches the Earth (Byrne et al., 2010). When a CME propagates away from the Sun, it interacts with the interplanetary medium and can be strongly transformed. Therefore, CMEs in interplanetary space are called interplanetary CMEs, or ICMEs. ICMEs are primarily magnetic clouds, the regions of high magnetic field intensities in the form of giant flux ropes (Burlaga et al., 1981, 1982). High magnetic field intensities in ICMEs can cause geomagnetic storms on Earth[3] if the magnetic fields have strong southward components (Gonzalez et al., 1994).

When an interplanetary CME (ICME) propagates at a speed faster than the ambient upstream solar wind by more than the upstream fast wave (magnetosonic) speed, a shock wave is formed in front of the ICME. (This is analogous to a jet fighter plane flying faster than the sound speed. A shock forms upstream of the plane wings. This shock can propagate down through the atmosphere and be heard and felt as a "sonic boom" by people on Earth). The properties of the solar wind are significantly altered between the shock wave and the ejected material. This region is called the sheath. The shock serves to compress the solar wind plasmas and magnetic fields, slow plasma flow to submagnetosonic speeds in the frame of the ICME/piston, and convert bulk kinetic energy into heat (Kennel et al., 1985; Tsurutani et al., 2011**).** The higher magnetic field intensities in the sheaths can cause intense magnetic storms, similar to ICMEs (Meng et al., 2019).

CMEs are closely connected to intense *solar flares,* or electromagnetic radiation bursts[4]. As far as we know, a significant fraction of X-class flares[5] are accompanied with CME releases and conversely, extremely fast CMEs (solar wind velocity $V_{sw} > 2,000$ km/s) have been associated with X-class solar flares. Yashiro et al. (2005) reported that the CME association rate for the flares in X1-X2 classes are 82-91%, and above X3 level all flares had associated CMEs. Less powerful flares (M- and C-class flares) are not always accompanied by CMEs (Burlaga et al., 1981; Klein and Burlaga, 1982). Yashiro et al. (2005) reported a ~20% CME association rate for C-class flares, and ~49% CME association rate for M-class flares. Both flares and CMEs are thought to be the result of magnetic field reconfiguration and reconnection at the Sun. The annihilated field energy goes into the flare electromagnetic energy, acceleration of charged particles and the release and kinetic energy of the CME. Although CMEs and solar flares are typically strongly related, there are exceptions. Yashiro et al. (2005) reported 4 X1-class flares (out of 49 X-class flares) without CMEs. Furthermore, some CMEs are related to "disappearing filaments" and some have had no obviously related flare/disappearing filament (Tsurutani et al., 1988; Tang et al., 1989; Tang and Tsurutani, 1990; Kamide and Kusano, 2015; Lakhina and Tsurutani, 2018).

---

[3] The list of near-Earth ICMEs for the period 1996-2022 could be found at
https://izw1.caltech.edu/ACE/ASC/DATA/level3/icmetable2.htm

[4] Curious readers may enjoy the Grand Archive of flare and CME cartoons as well as the list of relevant references at https://www.astro.gla.ac.uk/cartoons/index.html

[5] The standard flare classification is based on a peak flux intensity (W/m$^2$) in 1–8 A X-ray band. The three largest classes are X, M, C. The scale is logarithmic: a X1-class flare is 10 times more intense than a M1-class flare and a M1-class flare is 10 times more intense than a C1-class. Classification of X flares goes beyond X9 to X11, X28 (on November 4, 2003, the Sun's strongest X-ray flare on record), and to X100, X1000 and up for superflares from solar-like stars.

**Space Weather: from solar origins to risks and hazards evolving in time**

It is thought that around 10-20% of the energy stored in the magnetic field associated with a group of sunspots could be released as electromagnetic energy of flares, including UV to FUV, soft X-ray emissions and radio emissions (Leka and Graham, 2018). Different parts of the electromagnetic spectrum produce different Space Weather effects; see Section 4.4 for the description of different ionospheric effects.

Reconnection, solar flares and CMEs are associated with *Solar Energetic Particles (SEP)*. SEP are an increased flux of energized protons (up to GeV energies), heavy ions, and relativistic electrons, either coming directly from near-reconnection locations in the solar corona, or from CME-driven shock waves. Currently, the two main means of SEPs reaching Earth are considered to be: (1) interplanetary magnetic field lines traced backwards from Earth that connect to the flare site in the lower corona; (2) interplanetary magnetic field lines directly connected from Earth to the shock waves formed by propagation of the CMEs through interplanetary space. The first type of SEPs is called 'impulsive SEP events' and the second type is called 'gradual SEP events' because of the temporal profile of the particle radiation events observed near Earth (e. g., Desai and Giacalone, 2016; Tsurutani et al., 2009). The particle radiation storm from a gradual SEP event could last a few days or longer depending upon the transit time of the shock in front of the CME that propagates from the Sun to the Earth and beyond. Gradual SEPs are believed to be more important for Space Weather because impulsive SEPs are shorter in duration and induce less intense radiation levels**.** For more details concerning flares, CMEs and energetic particle acceleration, we refer the reader to recent review by Tsurutani et al. (2022a).

For the sake of completeness, we note that there are other sources of energetic particles, including Galactic Cosmic Rays (GCRs), particle acceleration at planetary magnetospheric bow shocks, CIR shocks, and the heliospheric termination shock (Reames, 1999). Aside from SEPs, GCRs are also important for Space Weather, especially when estimating the radiation danger for human flight (Lockwood and Hapgood, 2007). The level of GCRs changes normally in antiphase with the solar cycle, as the varying interplanetary magnetic field strength modulates GCRs entering the heliosphere. On the Earth's surface, modulation of GCRs by ICMEs cause "Forbush decreases", rapid decreases in GCR intensities due to particle deflection from the intense ICME magnetic fields.

SEPs cause particle *radiation storms* in the near-Earth environment and are responsible for numerous Space-Weather effects: Single Event Effects (SEE) and radiation impacts on avionics in general, increased ionization levels and variations of ionospheric densities in polar regions leading to radio wave absorption, increased radiation risks for astronauts and airplane crew members and passengers on polar flights, and Ground Level Enhancement (GLE) radiation at the Earth's surface (see also Sections 4.2, 4.3, 4.4 for specific Space Weather effects of SEPs.)

Prediction of radiation storm intensities is an important and not completely solved problem in the Space Weather community. It is impossible to predict SEP fluxes at present - on any meaningful timescale. This is partly due to yet unpredictable particle acceleration efficiencies and partly due to the complex geometry/magnetic connectivity issues. Energetic particles are guided by the interplanetary magnetic field lines. If the energetic particles accelerated at the flare site are released onto magnetic fields that connect to the Earth, those particles will be detected in the near-Earth environment. Particles accelerated at ICME shocks will reach the Earth if the field lines from that acceleration regions connect to the Earth. A complicating factor in the latter mechanism is that a CME shock has different particle acceleration properties at different locations along its surface (Kennel et al. 1984, 1985; Tsurutani and Lin, 1985).

**Space Weather: from solar origins to risks and hazards evolving in time**

Since solar flares (in particular the most powerful ones) are easy to detect using X-ray detectors onboard Earth orbiting satellites, and also because of the strong statistical links between energetic flares and large CMEs/SEPs, forecasting of solar flares is becoming a growing branch of research in Space Weather (Leka and Barnes, 2018). With the accumulation of new data from imagers, solar flare forecasting is beginning to use modern methods of statistical analyses, including machine learning (ML) and artificial intelligence (AI) techniques (Jiao et al., 2020).

Solar eruptive processes are linked to the most dangerous and disruptive effects/impacts and cause development of CME-driven *geomagnetic storms, radiation storms, substorms, and ionospheric effects.* Statistically, the most energetic eruptive phenomena tend to occur at solar maximum and slightly thereafter (Chen, 2011).

*Variability of solar irradiance* over the solar cycle and over solar rotation in the extreme ultraviolet (EUV, 30-120 nm) and far ultraviolet (FUV, 120-200 nm) spectral bands is one of the major factors defining the variability of ionospheric ionization and thermospheric neutral densities[6]. Thus, this is an important Space Weather variable (Woods and Rottman, 2002). FUV/EUV variability closely follows what is known as the International Sunspot Number (ISN). Statistical studies reveal a close correlation between ISN, solar radiance in the FUV/EUV range and solar radio flux at 2.8 GHz (10.7 cm wavelength), known as the F10.7 index[7] (e.g., Lukianova and Mursula, 2011). The underlying dynamics are thought to be complex and different for different spectral lines, showing the dependence from multiple processes in the photosphere, chromosphere and lower coronal (Floyd et al., 2005).

## 3. Geomagnetic Storms, Substorms and HILDCAA intervals

Since many Space Weather effects are related to the occurrence of *geomagnetic storms and substorms,* it is instructive to consider these processes in detail.

### 3.1. Geomagnetic storms

The interplanetary structures responsible for the formation of intense and large magnetic storms have been shown to be magnetic clouds (MCs, one of three parts of an ICME) and the upstream shocked sheath magnetic fields. CIR structures are responsible for milder geomagnetic storms. When the southward directed magnetic fields of the CIR, ICME or the sheath upstream of ICME interact with the terrestrial magnetic field, they cause an increased influx of energy into the magnetosphere. Solar wind energy is transported into the magnetosphere through magnetic reconnection at the dayside, the so called 'merging' of the solar wind magnetic fields with magnetospheric magnetic fields (Dungey, 1961). Magnetic field lines that are connected by one end to the solar wind and by another end to the Earth's ionosphere are dragged from the dayside to the nightside, causing accumulation of magnetic energy in the tail. After some threshold is reached, magnetic energy in the tail is released in magnetic reconnection and it causes the reversed plasma motion from nightside to the dayside. The whole process of plasma motion is called a global convection process. In the polar and mid-latitude ionospheres, this process creates a characteristic two-cell pattern of plasma motion called the Dungey convection cycle (Axford and Hines, 1961; Dungey, 1961).

---

[6] Only EUV radiation can ionize primary thermospheric species such as O, $O_2$ and $N_2$ (ionization energy thresholds are ~ 91, 103 and 80 nm respectively). An important thermospheric compound NO could be ionized by FUV Lyman-alpha radiation, contributing to the major ionization source at ~ 50-90 km, so-called D-layer of the ionosphere.

[7] The current sources of International Sunspot Number and F10.7 are the Sunspot Index Data Centre at Royal Observatory of Belgium for ISN (https://www.sidc.be/silso/datafiles) and National Research Council Canada/Natural Resources Canada (https://www.spaceweather.gc.ca/forecast-prevision/solar-solaire/solarflux/sx-en.php).

The increased energy input during geomagnetic storms is related to the opening of the magnetosphere to the interplanetary magnetic field, B. (From here and below, we assume Geocentric Solar Magnetospheric, GSM coordinate system for the magnetic field). For the interplanetary magnetic field $B_z$ oriented in the same direction as the Earth's dayside magnetopause magnetic field lines, the magnetosphere is said to be 'closed' with the minimal energy input from the solar wind (Tsurutani et al., 1992; Du et al. 2008). In the case where $B_z$ is oriented in the opposite direction to the Earth's magnetic dipole and magnetic reconnection occurs, the magnetosphere is said to be 'open'.

Although the interplanetary magnetic field $B_z$ and solar wind velocity $V_{sw}$ are the main factors contributing to the energy input into the Earth's magnetosphere, they are not the only solar wind parameters discussed in the literature (see e.g., Newell et al., 2007 and references therein). But in general, $V_{sw}$ x $B_{z,south}$ is the main contributor to the energy inflow.

When there is a prolonged interval of southward $B_z$ in the solar wind (a few hours), there is the formation of a torus-like region of enhanced fluxes of $H^+$, $O^+$ and electrons with energies ~10-300 keV. These particles which are initially injected into the nightside magnetosphere by Dungey's convection, drift around Earth under the combined effect of magnetic field curvature and magnetic field gradient drifts, forming a giant 'ring current' flowing in the space around the Earth (Dessler and Parker 1959; Gonzalez et al., 1994; Daglis, 2001).

The ring current creates a disturbance in the magnetic field strength (diamagnetic effect) that is detected at the Earth's surface. The disturbance is relatively small, and in most cases, it is less than 300 nT, or less than 1% of the Earth's magnetic field at the surface. However, the disturbance field has a characteristic time dependence. It covers ground-based magnetometer stations in a wide range of longitudes, lasting a few hours and then recovering more slowly over another half day to days. The ring current disturbance is measured at low geomagnetic latitudes (20°-30°) where the effect of auroral currents and equatorial electrojet are negligible and is called the disturbance storm-time or Dst index (now a one-minute SYM-H index is available). When the Dst index falls below -50 nT, it signals the occurrence of a geomagnetic storms (Gonzalez et al., 1994; Lakhina and Tsurutani, 2018).

The increase in magnetospheric convection during geomagnetic storms causes the increase of energy input into the coupled thermosphere - ionosphere system. Ionospheric plasma interacts with neutrals through collisions, therefore ionospheric currents generate a substantial amount of Joule heating that directly affects thermospheric (neutrals) uplift and satellite drag effects (See Section 4.5). Enhanced convection is also responsible for the transport of electrons and ions from the magnetospheric tail region toward the Earth and the inner magnetosphere, as well as electron and ion energization by conservation of the adiabatic invariants, and creation of the ring current. The interval of large southward $B_z$ component, when the strong ring current builds up, is the 'storm main phase'. It is often followed by an interval of northward $B_z$, making the magnetosphere more closed with minimal energy input. The stored ring current particles are lost through several different mechanisms (Kozyra et al. 1997). The storm phase in which the Dst index recovers is called 'storm recovery phase' and normally lasts a half day or more. The entire process could take a day or two. 10

**Space Weather: from solar origins to risks and hazards evolving in time**

In summary, a geomagnetic storm is a period of increased geomagnetic activity characterized by enhanced energy input into the coupled magnetosphere-ionosphere-thermosphere system, controlled to large extent by extended intervals of the southward component of the interplanetary magnetic field, high solar wind velocities (Gonzalez and Tsurutani, 1987; Echer et al., 2008; Lakhina and Tsurutani, 2016), and by other parameters as solar wind density and temperature (Newell et al., 2007).

The main features of magnetospheric dynamics and plasma circulation during geomagnetic storms can be reproduced by modern global 3D codes. These codes combine multiple modules to represent different regions and different physical processes of the coupled magnetosphere-ionosphere system (De Zeeuw et al., 2004; Toth et al., 2005; Fok et al., 2006; Moore et al., 2008).

Figure 4 shows an example output of a global code and demonstrates the structure of all major current systems in the magnetosphere, including the ring current. The global codes are mature enough to be used for forecasting geomagnetic activity at NOAA's Space Weather Prediction Center (SWPC) (https://www.swpc.noaa.gov/products/geospace-magnetosphere-movies). Despite the progress in development, many important issues remain unresolved. To name a few, it is not understood how to properly describe the ionospheric source(s) of plasma in global codes (so-called ionospheric outflow), how to describe and include multiple wave-particle interactions, and what is the role of kinetic effects on the global structure and dynamics of the magnetosphere.

## 3.2. Geomagnetic substorms and HILDCAA intervals

Another type of geomagnetic activity is a *geomagnetic substorm*. There can be many substorms within a magnetic storm or they can occur independently as isolated events. There are a variety of different models for the cause of substorms. In fact, all may be correct if applied for different events. We will describe below one commonly used model of isolated substorms. For an isolated substorm, the energy from the solar wind is first transported from the dayside to the nightside magnetosphere or magnetotail through the dayside magnetic reconnection process. In the tail, the energy is stored in the form of magnetic energy with a very stretched tail configuration and large values of magnetic field B (substorm growth phase). Since the process of energy accumulation is not infinitive, at some point the magnetosphere loses its stability and the energy is released in non-stationary tail reconnection (substorm onset). After the substorm onset, the stored magnetic energy is transformed into kinetic energy of plasma (note the analogy with CME release, Section 2), appearance of auroras and intense ionospheric currents (substorm expansion phase). Substorms cause additional particle transport and energization, that is, particle substorm injections into the ring current plasma. The last step is the return of the system to the initial state (substorm recovery phase).

High-Intensity, Long-Duration, Continuous AE Activity (HILDCAA: Tsurutani and Gonzalez, 1987) intervals are by definition intervals of enhanced auroral activity and are a combination of cyclical substorms and DP2 (Nishida, 1968) events (Tsurutani et al. 1995; 2004a). The most intense substorms occur during the solar cycle declining phase when there are large coronal holes at the Sun that emit HSSs of solar wind plasma. HILDCAA activity is produced by magnetic reconnection associated with embedded Alfvén waves within the HSSs. HILDCAAs and enhanced electromagnetic chorus waves produce enhanced fluxes of relativistic MeV "killer" electrons in the magnetosphere (Hajra et al., 2015; Hajra and Tsurutani, 2018).

**Space Weather: from solar origins to risks and hazards evolving in time**

It is important to understand that substorms are sometimes parts of geomagnetic storms and sometimes not. Substorms are related to the buildup/release of magnetic energy, and it is advantageous to consider them independently from the buildup of kinetic energy in the ring current during geomagnetic storm. The buildup of kinetic energy and induced changes of B-field at nightside may alter stability properties of the tail. It has been reported that storm-time substorms and isolated substorms have different characteristics (e.g., Pulkkinen et al., 2007; Partamies et al., 2013). There are other modes of magnetospheric behavior besides cyclical substorms, such as steady magnetospheric convection events (convection bays: Sergeev et al., 1996) and recurrent internally driven substorms (Keiling et al., 2022).

Examples of isolated substorms can be found in Tsurutani and Meng (1972) and in HILDCAAs (Tsurutani and Gonzalez, 1987). For both cases, where there is insufficient ring current activity to indicate the presence of a magnetic storm (Dst/SYM-H > -50 nT) . On the other hand, it has been shown that in some MC-induced storms, there are no substorms (Tsurutani et al., 2004c). This has been speculated as being caused by the very smooth rotations of the IMF Bz. Are these the Sergeev et al. (1996) convection bays but giant in size? It is not clear at the present time.

The magnetosphere often transitions between different modes, and possibly exhibits a memory, where the present mode is defined by the prehistory of both the solar wind input and magnetospheric/magnetotail dynamics. At present, it is difficult for scientists to accurately predict what type of magnetospheric activity will be induced by a given solar wind/interplanetary magnetic field input. This is currently an active area of research.

## 4. Major Space Weather Effects and Impacts

We have briefly reviewed some of the Space Weather processes shown in Figure 2. The system is very complex, and many Space Weather effects and impacts could be linked to multiple solar origins. Among the most important Space Weather effects/impacts are:

(1) Geomagnetically Induced Currents (GICs); (2) radiation and plasma effects/impacts related to surface charging and arcing, deep dielectric charging, avionics and s/c electronics; (3) effects/impacts of radiation related to human health, (4) ionospheric effects/impacts related to navigation and communication; (5) thermosphere effects and satellite drag. The above divisions are not unique. Since Space Weather science has many facets, there are multiple ways to organize these classifications. The detailed description of each Space Weather impact is a separate branch of Space Weather science; therefore, below we only briefly present the most important Space Weather effects/impacts.

## 4.1. Geomagnetically Induced Currents (GICs)

As mentioned above, the first encounters of GIC effects on human technologies took place in the middle of the 19[th] century with the widespread development of telegraph systems. Now, GICs are recognized as a one of the main Space Weather threats. The underlying physics could be explained by Faraday's law: changes in the large currents systems (e.g., electrojet) cause a changing magnetic field that induce electric fields (Lakhina et al. 2020). The electric fields drive strong currents on the ground if the conductivity is large enough. These induced currents flow near the Earth's surface.

Sometimes the currents can get into transmission cables and power lines and cause strong GIC effects. Any technological system that has good electrical conductors as an essential design element and also by design electrically connected to the Earth is potentially vulnerable to the effect of GICs. The best-known examples are power grid outages and transformers failures, with more exotic cases including railway system failures, pipeline monitoring system failures and impacts on undersea telecommunication traffic cables (Lanzerotti, 1992; Boteler and Jansen van Beek, 1999; Viljanen et al. 1999; Wik et al., 2009; Eroshenko et al., 2010; Hapgood, 2018; Ngwira and Pulkkinen, 2018).

Variations of the geomagnetic field and resultant GICs are due to the reconfiguration of the giant current loops circulating in the near-Earth environment. These currents create a magnetic field and shape the magnetosphere. For example, the magnetic field of the magnetopause current (Chapman-Ferraro current) defines the shape of the magnetopause, and the cross-tail current (and current loops) defines the shape of the geomagnetic tail. When the global current system reconfigures itself as a result of dynamic changes in the magnetic field/plasma conditions of the solar wind (e.g., geomagnetic storms) and/or changes in the balance of stored energy inside the magnetosphere (e. g., magnetospheric substorms), the ionospheric part of the current loop also changes, producing strong and fast variations of the magnetic field at the Earth's surface. The main drivers of GICs are thought to be:

(i) Substorm ionospheric currents and variations in the auroral electrojet (Pulkkinen et al., 2003; Viljanen et al., 2006; Tsurutani and Hajra, 2021). The ionospheric currents can be as large as $10^6$ Amperes and located only ~100 km above the Earth's surface;

(ii) Rapid reconfiguration of dayside Chapman-Ferraro currents related to shock wave/sheath impact and possible interactions with the equatorial electrojet (Carter et al., 2015). Storm commencement due to shock impact is believed to be the main driver of low-latitude GICs (Gaunt and Coetzee, 2007).

(iii) Other sources of strong geomagnetic variations include geomagnetic field pulsations, possibly associated Kelvin-Helmholtz instabilities, and possibly related ring current reconfigurations (Buzulukova et al., 2018b).

Recent studies show that intense GICs are correlated with the occurrence of intense substorms (Tsurutani and Hajra, 2021). However, it is very difficult at present to predict the occurrence of an intense substorm and related intense GICs at a particular location because GICs could be very localized (Ngwira et al., 2015; Pulkkinen et al., 2015; Tsurutani and Hajra, 2021), and likely related to transient structures, bursty bulk flows or depolarization flux bundles or both, sub-units of geomagnetic substorms (Kepko et al., 2015). Global models of the magnetosphere are improving each year, but at the current state of knowledge, it is difficult to reproduce intense GIC structures as observed in the data. A detailed summary of the challenges of predicting and understanding GICs is presented in Thompson et al. (2010), Pulkkinen et al. (2017) and Lakhina et al. (2020).

## 4.2. Particle radiation and plasma effects on satellites

Exposure to radiation from a variety of sources including energetic electrons, protons, heavy ions and neutrons degrades satellite systems. When properly designed, electronic and avionic components of s/c systems should meet certain aerospace electronics standards to be resilient to radiation exposure in a particular orbit, e.g., LEO, Medium Earth Orbit (MEO), or Geosynchronous

**Space Weather: from solar origins to risks and hazards evolving in time**

orbit (GEO). High-energy particles passing through microelectronic devices undergo a chain of complex interactions with the material of the device, depositing the energy and creating charge. There are a few main classes of radiation effects that are taken into account when designing satellite systems: *single event effects (SEEs), total ionizing dose, and displacement damage dose. Surface charging and arcing*, as well as *deep dielectric charging* are usually considered separately (Baker et al. 2018).

*SEEs* are broad class of anomalies related to a single passage of an energetic particle, the production of an electron-hole pair[8] in a silicon substrate being probably the most common case (Baumann, 2005). These can include pulses in logic or bitflips in memory cells. They are mostly non-destructive and transient, but could potentially become an overload during a radiation storm. Some SEEs may require a power reset, and some are considered potentially destructive. *Total ionizing dose effects* are related to the amount of energy deposited by energetic particles, and displacement damage dose refers to the creation of defects in semiconductor lattices (displacement of atoms from their lattice sites caused by passage of the energetic particle). Total ionizing doses and displacement doses are called cumulative radiation effects, and should be estimated for a given orbit. More information on radiation effects could be found in the NASA GSFC Radiation Effects and Analysis web-site on https://radhome.gsfc.nasa.gov/top.htm, NASA Applied Space Environments web-site on https://www.nasa.gov/nase, and ESA's SPace ENVironment Information System (SPENVIS) on https://www.spenvis.oma.be/. See also Barnaby and Mariella (2013) and Tsurutani et al. (2000). The latter was a NASA white paper written by an international group of particle scientists and engineers for the protection of the Parker Solar Probe mission.

There are multiple Space Weather processes that cause radiation damage. The main sources are: GCRs, SEPs, and trapped radiation from the electron and proton radiation belts. The GCR background is well-known. The high-intensity proton inner radiation belt fluxes and spectra are relatively well known and vary slowly with solar cycle except SEPs effects (Selesnick and Albert, 2019; Lozinski et al., 2021). Therefore, the most important sources of radiation for Space Weather effects are solar energetic protons and ions (SEPs/radiation storms), and MeV electrons of the inner/outer radiation belts (substorms/CIR storms/HILDCAA events). A review of radiation belt processes can be found in Baker et al. (2018).

Surface charging/arcing effects and deep dielectric charging remain serious Space Weather threats to satellites. Surface charging/arcing effects refer to an accumulation of electric charge on parts of s/c due to the interaction with the ambient energetic plasma. An enhanced level of electron fluxes in the energy range 100 eV-50 keV is known to be a hazardous environment. A significant potential difference between different (not electrically connected) parts of s/c[9] (up to a few kV) could build up over a relatively short period of time (minutes). This could cause s/c anomalies and loss of satellite control. A potential difference of a few kV can cause arcing effects, solar arrays being one of the most vulnerable parts. LEO satellites passing through the auroral zone, and satellites in GEO, are believed to be the most vulnerable to surface charging effects. The main Space Weather processes responsible for surface charging are thought to be geomagnetic substorms or substorm-like transients (Section 3.2), which cause enhanced fluxes of 1-50 keV electrons in the GEO region, or enhanced fluxes of precipitating electrons in the auroral zone (DeForest, 1972; Spence et al., 1993; Ganushkina et al., 2021). More details about surface charging can be found in Ferguson (2018).

---

[8] In silicon, one electron-hole pair is produced for every 3.6 eV of energy lost by the ion. Therefore, a single 1 MeV ion could produce up to ~ 278,000 electron-hole pairs. See Bauman, 2005 for the details.

[9] To reduce the risk of electrical discharges between different surface parts, a s/c needs to be designed and built so that all surface parts are electrically interconnected. Although efforts have been made to make the s/c surfaces as conductive as possible, the spacecraft area still contains insulating materials, for example solar cell cover glass (Puthanveettil et al., 2014).

Deep dielectric charging refers to the accumulation of electric charge and intense electric field build-up, caused by more energetic particles, usually electrons with energies up to MeVs, but also by ions of MeV energies, because the latter could amplify charging (Lai et al., 2018). If the electric field reaches a critical value $10^6$ - $10^8$ V/m depending on the material, there is a high probability for dielectric discharges to occur. The discharges can cause s/c anomalies, with potential damage to some important s/c systems, e. g., printed circuit boards. Since the high energy electron flux is far less than that of 1-50 keV electrons, the accumulation of sufficient charge for a discharge could take days or even weeks, therefore showing a delay time from the onset of the charging. Statistically, most of the internal charging/discharging effects have been shown to occur during the declining/minimum phases of the solar cycle (Wrenn et al., 2002; Lohmeyer and Cahoy, 2011). These results are consistent with findings that outer MeV electron radiation belt tends to intensify during the declining phase (Miyoshi and Katoka, 2011). The main Space Weather processes responsible for the deep dielectric charging effects are geomagnetic substorms, or substorm-like events (HILDCAAs) that cause outer MeV radiation belt fluxes to intensify (Hajra et al. 2015).

More information about s/c charging could be found in the NASA Technical Handbooks NASA-HDBK-4002A w/CHANGE 1, NASA-HDBK-4006 as well as in Ferguson and Hillard, 2003; Garrett and Whittlesey, 2012; Ferguson, 2018; Zheng et al., 2019.

### 4.3. Radiation effects related to human health

Studies of Space Weather radiation effects on human health could be divided into two broad groups: (1) radiation field in space and assessment of the health risks for astronauts aboard the ISS, or crewed missions to Mars and Moon; (2) radiation field at aviation altitudes and assessment of health risks for aircrew, high-altitude pilots and frequent travelers.

Two major sources of radiation risks for humans in space are: major SEP events and GCRs. GCR ions are more energetic than SEP ions and electrons with a broad spectrum of GeV energies and higher. GCR ions have an isotropic distribution whose flux variations during the solar cycle are well-known. GCR particle fluences are therefore predictable and could be calculated for a given orbit; however, due to the very high energies of GCRs it is almost impossible to construct an efficient shielding outside of the Earth's magnetic field and atmosphere.

SEPs typically have a spectrum with energies of up to hundreds of MeV. The spectrum is difficult to predict due to the underlying complexity of the formation mechanisms (see Tsurutani et al., 2020). The fluxes for major SEPs are much higher than compared to background GCRs. The threat for astronauts from SEPs is not hypothetical. Estimations show that the August 4, 1972 event that occurred a few months before NASA's Apollo 17 landed would have been lethal for astronauts walking on the lunar surface (Lockwood and Hapgood, 2007). However, for long-duration missions (e.g., to Mars) chronic exposure to GCRs will likely be more challenging (Schwadron et al., 2014; Dobynde et al., 2021). Due to a very different radiation profile, chronic exposure for GCRs and acute burns for SEPs, mitigation approaches are likely to be very different (Narici et al., 2018; Hapgood et al., 2019).

**Space Weather: from solar origins to risks and hazards evolving in time**

Radiation dose and its effects on avionics at aviation altitudes has been studied with environment monitors onboard supersonic Concorde flights (Dyer et al., 1989). Awareness of natural radiation at aviation altitudes as a health hazard began to increase in the early 1990s (Band et al., 1990; Barish, 1990). In 1994, the US Federal Aviation Administration (FAA) formally recognized the associated health risks (Friedberg and Copeland, 2003). This triggered the need for quantification and characterization of the global radiation environment at aviation altitudes (Friedberg and Copeland, 2011; Tobiska et al., 2016). This is an area of ongoing research activity, both for modeling and measurements.

The radiation field of concern at aviation altitudes mainly originates from SEPs, GCRs and relativistic magnetospheric electrons. SEPs and GCRs with energies more than 400 MeV are a hazard for low and middle magnetic latitudes. Even for quiet Space Weather conditions, the radiation field at aviation altitudes is more than an order of magnitude higher than at the ground level. After the great Halloween geomagnetic storms of 2003, the Space Weather community realized that the radiation index designed for geostationary orbit (particle flux with energy 10 MeV and above) was insufficient for aviation since a major part of the 10 - 100 MeV radiation is shielded by the Earth's atmosphere. A special index for the aviation radiation environment has been introduced, the D-index (Meier and Matthia, 2014). Extreme environments are still difficult to monitor during real-time flights, which is why the radiation field estimations are made from historical GLE observations on the ground. Dyer et al., 2018 provided a detailed analysis based on the 23 February 1956 extreme GLE. Appropriate mitigation measures, such as lowering aircraft altitude with minimal additional fuel consumption, have been shown to reduce radiation exposure during major events by up to 40% (Matthia et al., 2015).

Magnetospheric relativistic 1 to 10 MeV electrons, such as GCRs and SEPs, do not reach airplane altitudes. However, they form a "shower" by cascading into other particles like muons, gamma ray photons, electrons and positrons (Tsurutani et al., 2016). This radiation has just recently been realized to be a hazard to airplane pilots and personnel (Tobiska et al., 2018).

In November 2019, a new service for global civil aviation has been launched under the auspices of the International Civil Aviation Organization (ICAO), to provide real-time and worldwide space weather updates for commercial and general aviation. This service aims to generate and share space weather advisories using data collected from dedicated space weather centers established by 17 countries (https://www.icao.int/Newsroom/Pages/New-global-aviation-space-weather-network-launched.aspx).

**4.4. Ionospheric Space Weather effects related to navigation and communication**

It has been just 125 years since the invention of the first radio wave transmitter and receiver, but it's hard to imagine our life without radio communication devices. What if an extreme Space Weather event disrupted all of these systems?

The range of radiocommunication frequencies extends from a few 10s of kHz for ground-wave propagation to a 10s of GHz for satellite communication frequencies[10]. The choice of frequencies depends on the systems and their implementations. However, there are common underlying physical processes that control the propagation of radio waves in the ionosphere, and therefore control the effects.

---

[10] Frequencies ~ 10 GHz and up are likely not very sensitive to space weather. See also https://www.alsa.mil/News/Article/2532178/true-impacts-of-space-weather-on-a-ground-force/ for some examples.

(1) The ionosphere has a layered structure, with a few layers of increased ionization. Each layer (D, E, F) is created by a particular ionizing source of photons from the Sun (UV, EUV, X-ray), and SEPs and auroral precipitation, thus naturally forming the dependence on Space Weather processes. Radio waves can only propagate at a frequency higher than the plasma frequency, which depends on the square root of plasma number density (electron number density). The plasma frequency for the Earth's ionosphere generally varies in the range of 2-20 MHz, defining the frequency range for HF communications. Therefore, any change in electron number density in the ionosphere caused by changes in the ionizing sources will have a significant effect on wave propagation in the radio HF range.

(2) In the context of GNSS signals, the important quantity is an ionospheric delay, the error in estimating the Euclidian distance between the receiver and the s/c due to the group speed/phase delay of the wave packet in the ionosphere. More simply stated, the ionospheric delay occurs because the ionosphere affects the propagation path and speed of radio waves, making the former typically longer (due to refraction) and the latter less than in a vacuum. The ionospheric delay is related to the plasma density between the receiver and the s/c. The total electron column density is called the slant Total Electron Content (TEC), which in turn could be recalculated into vertical TEC, or VTEC. Global VTEC maps are needed to calculate ionospheric delay and GNSS signal propagation errors for users with single frequency receivers. Dual-frequency GNSS receiver data are used to calculate and remove the contribution of ionospheric delay using two of the broadcast frequencies from the GNSS satellites. When dual-frequency reception is not available, single-frequency GNSS receivers rely on ionospheric models and data from the dual-frequency ground reference station network to estimate VTEC values. Any sudden and localized changes that are not reflected in the model or in the VTEC measurements will cause errors for users of single frequency receivers. An example of a system that currently supports single-frequency user receivers is the US GPS-based Wide Area Augmentation System (WAAS), designed to assist civil aviation users (Sparks et al., 2021). This type of Space Weather hazard has potential to decrease in the future, when users are able to use dual-frequency receivers more widely, with the advent of a GPS constellation that transmits a second frequency.

(3) Turbulence in the ionosphere can cause inhomogeneities in ionospheric plasma density that refract/diffract the incoming radio signals and produce interference patterns (Ghobadi et al., 2020)**.** Inhomogeneities at scale sizes of the order of a Fresnel zone (10s-100s m) are effective[11]. In the context of GNSS signals, these patterns introduce rapid fluctuations in the wave phase and the amplitude, known as ionospheric scintillation. If the scintillations are large enough, they prevent the receiver from maintaining lock on an incoming signal, i.e., disrupting tracking of the signal by the receiver. Since scintillations are caused by small-scale fluctuations in ionospheric densities, it remains challenging to model and predict the effects of this phenomenon.

(4) The coupling between electrons and neutral plasma at low altitudes is responsible for radio wave absorption in the D-layer (60-90 km) of the ionosphere. If the plasma density suddenly increases in this region where the electron-neutral collision frequency is higher than the wave frequency, then the waves propagating through these regions will be absorbed. This effect is called a

---

[11] The Fresnel zone defines the distance scale over which destructive interference can occur.

'sudden ionospheric disturbance' or SID, and is responsible for radio blackouts within the HF range caused by X-ray flares.

(5) Solar radio bursts (SRBs) have the potential to directly interfere with radar systems[12] in a wide range of frequencies (Hey, 1946; Knipp et al., 2016; Marqué et al., 2018) and GNSS signals in the frequency range that falls within the range of the GNSS transmitters, 1-2 GHz (Klobuchar et al., 1999; Cerutti et al., 2008). This effect is less common than ionospheric scintillations, but could be profound for particularly large events, thus defining SRBs as an important process in Space Weather (Bala et al., 2002; Gary and Bastian, 2021). Solar radio bursts can last from a few hours to a few days. It is important to define what is the radio flux threshold in the GNSS frequency band that will introduce significant errors and disturbances for a given system.[13] Recent studies show that this threshold may be lower than originally thought (Yue et al., 2018).

Some of the most important Space Weather processes that control ionospheric variability are the UV/FUV/EUV/soft X-ray part of the solar radiation that comes from both background processes (active regions) and eruptive processes (energetic solar flares), particles of SEPs, and geomagnetic storms and substorms. Studies of energetic flares demonstrate that the combined effect of UV/FUV/EUV/soft Xray radiation could increase the ionospheric photoionization rate within minutes, thus introducing rapid ionospheric variations (Tsurutani et al., 2005).

On the nightside ionosphere, magnetospheric precipitation of electrons with energies 0.1-100 keV (Newell et al., 2009; Miyoshi et al., 2015) is the main source of ionospheric ionization and it is closely related to storm/substorm activity. Geomagnetic storms/substorms inject energy into the magnetosphere and ionosphere. It creates plasma instabilities and particle precipitation in the coupled thermosphere-ionosphere system, and irregularities in the ionospheric density. Complexity of the coupled system therefore defines the response over a wide range of spatial and temporal scales.

## 4.5. Thermosphere effects and satellite drag

The ionosphere is closely coupled with the upper Earth atmosphere, or thermosphere (~90-900 km) via collisions with neutrals. One major Space Weather effect of concern is the expansion of the upper atmosphere during magnetic storms where there is atmospheric heating in the auroral and subauroral zones. Collisions with neutrals produce an atmospheric drag effect for s/c in LEO. The result is a gradual reduction of s/c altitude, and, in extreme cases such as during intense magnetic storms, a loss of s/c track. An example could be the geomagnetic storm of March 1989, when after 5 days ~ 1000 satellites became "lost" for a few days (Hapgood, 2018, Fig. 7).

The coupling between the thermosphere and ionosphere allows an exchange of momentum between the plasma and the neutral atmospheric species. As a result, global motion of the neutral component (e.g., winds) could drive plasma flows, and vice versa, strong plasma flows could drive neutral flows. This makes the variations of neutrals in the upper atmosphere closely coupled with Space Weather processes (Mannucci and Tsurutani, 2018; Fuller-Rowell et al., 2018; Crowley and Azeem, 2018; Hapgood, 2018). At the present time computer codes can model neutral-ion coupling, but there does not exist an adequate code for ion-neutral coupling for extreme storms like the Carrington event at present (Deng et al., 2018).

---

[12] The effects of solar radio bursts on radars were first presented in secret reports during World War II documenting widespread disturbances of British radar systems operating at 55-85 MHz on February 26-28, 1942 (Hey, 1946).

[13] Specification of the threshold for radio burst interference is receiver dependent.

One important parameter that controls atmospheric expansion is ionospheric Joule heating (Bates, 1974). Joule heating is the main channel for magnetospheric energy dissipation in the ionosphere: energy deposition into ionosphere due to magnetospheric energetic particle precipitation is small in comparison with energy deposition due to Joule heating (Wilson et al., 2006). But precipitation controls ionospheric conductance (Robinson et al., 1987), and hence the Joule heating pattern. To calculate Joule heating it is necessary to know global distributions of ionospheric conductance and electric fields. This problem is still not solved completely, and uncertainty for estimations of Joule heating rate translates into uncertainty for satellite tracking.

Another important uplifting mechanism for neutrals that has recently gained attention is the "equatorial plasma super-fountain" (Tsurutani et al., 2004b; 2008; Mannucci et al. 2005). The global convection electric field during storms is known to penetrate to low ionospheric latitudes in the equatorial region (Nishida, 1968a,b). This is called the Prompt Penetration Electric Field (PPEF). It is thought that PPEF arises from insufficient shielding of the global convection electric field by the ring current via Region II Birkeland currents flowing in/out of the ionosphere (Toffoletto et al., 2003). Under quiet conditions, the PPEF is weak but intensifies during the storm main phase. Estimates show that during strong geomagnetic storms the PPEF will cause the dayside upper atmosphere to be uplifted through ion-neutral drag in the near-equatorial regions (Lakhina and Tsurutani, 2017).

There is an important Space Weather process affecting neutral density variations that is not related in general to geomagnetic storms. Increased solar UV, FUV, and EUV radiation directly heats the thermosphere, therefore variabilities in these quantities will also cause atmospheric expansion and uplift. Increased levels of solar radiation are linked to the group of sunspots, active regions, rotating with the Sun and therefore lasting a few weeks. Estimates show that a moderate increase in EUV/FUV level could increase the thermospheric neutral density at 400 km and 850 km by 100% and 200%, correspondingly (Fuller-Rowell et al., 2018).

Space Weather effects on satellite tracking is one of risks that evolve in time and currently is increasing. This is related to a more general problem of collisions with space debris. It is currently a real challenge since there are thousands of satellites being launched into LEO (e.g., Starlink). A concern is that an extreme Space Weather event similar to what happened in the past would increase satellite drag unpredictably. This in turn would increase the risk of collisions with space debris/other satellites and start the Kessler effect or 'domino effect' when each subsequent collision increases chances for the next one (Kessler and Cour-Palais, 1978).

To summarize, the factors controlling satellite drag effects and influencing satellite orbit prediction and collision prevention are: (1) neutral winds as they affect s/c orbital velocity and also affect the coupling between neutrals and plasma in general; (2) neutral composition as it modulates the expansion by changing scale heights; (3) global plasma flows horizontally and vertically, because they translate into Joule heating rates and drive motions of neutrals (through ion-neutral collisions); (4) modulation of global heating rate by radiative nitric oxide (NO) cooling (e.g., Knipp et al., 2017); (5) variations in solar UV, FUV and EUV radiation related to Sun's active regions due to direct heating of the neutrals; (6) the Kessler effect in the case of a hypothetical extreme Space Weather event.

## 5. Discussion: Major challenges and the Worst-Case Scenario

Given the growing importance of the Space Weather effects for a technologically advanced society, it is important to understand and anticipate the associated risks and dangers. Space Weather regularly affects satellite communications and avionics, for example with partial degradation of GNSS services or planned gradual radiation damage to satellites. Society is developing mitigation techniques and in parallel a better understanding of the underlying physical processes. However, there is one very important area where we still have significant fundamental knowledge gaps. This area of concern relates to the prediction of extreme solar eruptive events and their interactions with Earth.

Solar eruptive phenomena result from the constantly changing solar magnetic field, plasma instabilities and finally from magnetic reconnection in the solar corona. However, at the present it is unclear how reconnection controls the ejection of CMEs (e.g., Gou et al., 2019; Zhu et al., 2020). As a result, it is very difficult, if not impossible, to predict whether a group of sunspots will cause an event on Earth of similar intensity to the Carrington event. One principal difficulty is predicting the direction of the CME magnetic field and if it remains in the same direction as it propagates from the Sun to the Earth (Tsurutani et al., 2020). To add complexity, it should be noted that the Carrington event was not accompanied by an intense radiation storm (Wolff et al., 2012). Solar flares, radiation storms and geomagnetic storms are related Space Weather phenomena, but can occur together or independently.

With growing data sets available, it seems possible that soon we will have statistical and/or AI-based models accurate enough to predict that an energetic flare will occur with some lead time so society can implement mitigation measures. Then the next step is to predict the occurrence of a radiation storm on Earth, International Space Station, the Moon or Mars, if the area of concern also includes human space flight missions. This is not easily obtainable in general, since there is not a simple cause-effect relationship between an intense flare and radiation storms. Currently it is difficult for physics-based models of the heliosphere to accurately predict the propagation of a CME through the interplanetary medium or make an accurate prognosis for SEPs (again see Tsurutani et al. 2020 for detailed discussions). We note that accurate prediction of CME trajectory, CME arrival time and CME properties (e.g., profiles of magnetic field near the Earth) is widely recognized as a strategic goal for the international space weather community. As noted by Riley et al. (2018), progress is slow. However, results from recent works that combine physics-based models, data assimilation and machine learning methods are cautiously optimistic (Amerstorfer et al., 2020; Kay et al., 2022; Alobaid et al., 2022).

McIntosh et al. (2015) showed examples where bursts of extreme solar activity occur as a changing pattern of solar magnetism with characteristic period 6-18 months. These bursts have been attributed to the existence of solar Rossby waves that are formed in the. solar 'tachocline', a region of solar interior that separates the differentially rotating convection zone from solidly rotating radiative zone (see the review by Dikpati and McIntosh, 2020). Recent advances in modeling and understanding of solar Rossby waves, coupled with advances in data assimilation techniques, promises to deliver space weather forecasting on a time scale from a few weeks to a few years (Dikpati and McIntosh et al., 2020). In addition to the Rossby waves hypothesis, there are other recent developments to understand solar cycle variability and quantify the likelihood of severe space weather events during different phases of solar cycle (Chapman et al., 2020; Leamon et al., 2022).

**Space Weather: from solar origins to risks and hazards evolving in time**

What would be the worst-case scenario? How strong would the greatest geomagnetic storm be? Radiation storm? Geomagnetically induced currents? How would this affect the avionics/satellite electronics? Telecommunications? GNSS services? Power grid infrastructure? These questions are becoming important as different countries begin to realize the threats and dangers associated with extreme Space Weather conditions. As stated in recently submitted report from the UN Expert Group on Space Weather (UN-EGSW)[14], 'improved international cooperation and coordination can lead to improved global resilience and preparedness in response to the adverse impacts of space weather.' Estimations made by the Royal Academy of Engineering in the UK show that an extreme space weather event will be difficult to deal with, but not catastrophic (Cannon, 2013; Hapgood et al., 2021). But we note that our current knowledge is based on limited statistics of extreme events that have happened in the past.

Could magnetic storms more extreme than the Carrington event occur in the future? The answer is yes, but a following question is how much larger? Using the maximum observed CME speed of 3,000 km/s, Tsurutani and Lakhina (2014) showed that it is theoretically possible to have a magnetic storm that is twice the intensity of the Carrington event. However, the authors mentioned a caveat, "if the magnetosphere does not saturate at $\beta=1.0$" ($\beta$ is the ratio of plasma thermal pressure to magnetic pressure).

In August, 1972 a solar active region launched several CMEs in the direction of the Earth. Although the CME that impacted the Earth was faster than the Carrington CME, the 1972 CME had northward IMFs so the resultant geomagnetic storm was small (Tsurutani et al., 1992). However, at the same time the AT&T L4 telecommunications cable line was down for an hour and the radiation flux was one of the highest during the space age (Lanzerotti, 1992). This was a case of extreme Space Weather effects (CME speed, radiation and GIC effects) with a low magnetic storm intensity. All of these effects were related but obviously not strongly correlated in intensities.

On July 23, 2012 the Sun released an extremely fast CME with a transit time to 1 au of 19 hours, slightly slower than the 17 hour 40 minutes of the Carrington event. Baker et al. (2013) estimated that this could have produced a magnetic storm of peak intensity Dst =-1182 nT. Luckily, the 2012 CME missed the Earth!

What can we learn from other stellar systems? Data from solar-type stars could help to accumulate necessary statistics on extreme events and estimate worst-case scenarios for our Sun. Maehara et al. (2012) studied "superflares" at solar-type stars. These superflares are 10 to 1,000 times more intense than the most intense flares recorded on our Sun. Extending these previous works, Okamoto et al. (2021) analyzed all of the Kepler primary mission data, applying the method of Notsu et al. (2019), and suggested that the Sun can cause superflares with energies of $\sim 7 \times 10^{33}$ erg ($\sim$X700-class flares) and $\sim 1 \times 10^{34}$ erg ($\sim$X1000-class flares) once every $\sim$3000 and $\sim$6000 years, respectively.

Takahashi, Mizuno, Shibata (2016) showed that the upper limit of the CME speed associated with X1000 and X100 superflares could be 9100 km/s and 6200 km/s, respectively (see Fig. 5 of Takahashi et al. 2016). Takahashi and Shibata (2017) further showed that these superflares are expected to produce massive CMEs (M $\sim$ 6 x $10^{17}$ g for X1000 flares) so that the deceleration

---

between the Sun and the Earth would be small. That means that giant sunspots at our Sun could possibly produce a superflare and a CME with potentially disastrous Space Weather consequences (Shibata et al., 2013).

Miyatake et al. (2013) and Usoskin et al. (2013) have studied tree ring and ice core data and have concluded that enormous radiation events or a series of events have occurred in 774-775 AD and 992-993 AD. If a similar event happens today, what would be the consequences? We know some of the answers, but they are incomplete. It is very important to identify the realistic worst case, along with the probability of them happening, and the potential impacts assessed. Then we need to make informed decisions on what level of risk we should take as a society, or as a business, and at what cost. It is our responsibility to understand and determine the variability and extremes of the natural environment and calculate their consequences.

## Funding

For N. B., this work has been partially supported by NASA grant 80NSSC19K0085. Portions of this research was performed at the Jet Propulsion Laboratory, California Institute of Technology under contract with NASA.

## Acknowledgements

N.B. thanks ISSI in Bern Switzerland for support of the International Team #523 "Imaging the Invisible: Unveiling the Global Structure of Earth's Dynamic Magnetosphere". B.T.T. thanks ISSI in Bern Switzerland for support of the International Team #455 on "Complex Systems Perspectives Pertaining to the Research of the Near-Earth Electromagnetic Environment". The authors thank P.F. Chen, R. Hajra, R. B. Horne, G.S. Lakhina, A. J. Mannucci, and K. Shibata for their reviews of the draft version of this paper. What was particularly useful was their expertise in specific detailed areas of Space Weather. Their comments were invaluable.

## Figures

(a) Drawing of auroral observations on the evening of September 2nd , 1859

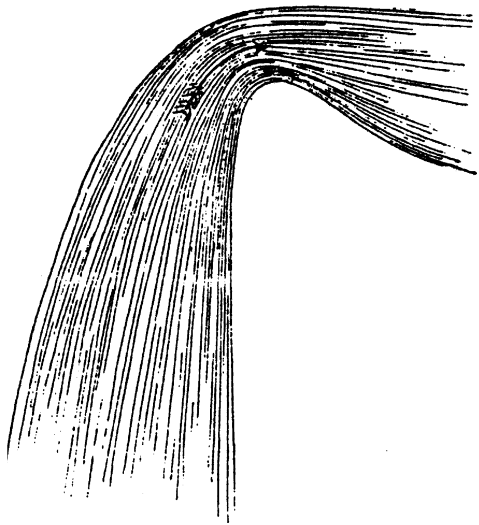

(b) Part of an auroral oval as seen from International Space Station

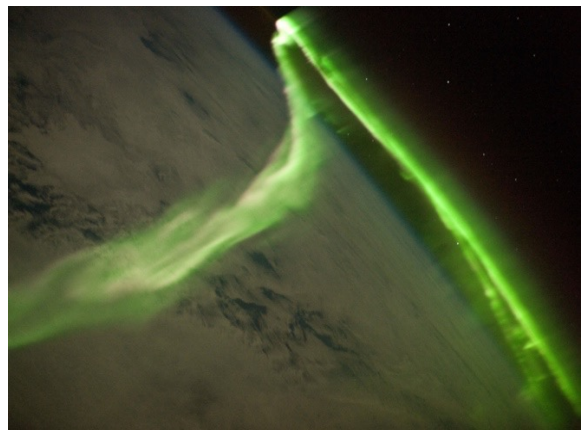

**Space Weather: from solar origins to risks and hazards evolving in time**

*Figure 1. (a) A picture of the auroral display on the evening of September 2$^{nd}$, 1859, Rome, N.Y., Lat. 43 deg 13 min designed by Mr. Edward Huntington. The picture was originally published in a "5$^{th}$ article on Great Auroral Exhibition" by Prof. Elias Loomis, in the American Journal of Sciences, 1859-1861. Credit: COSPAR and Advances in Space Research. Source: Shea and Smart, 2006. (b) Part of an auroral oval as seen from the International Space Station, taken during a geomagnetic storm in 2010. Credit: NASA. Source:*
*https://commons.wikimedia.org/wiki/File:Aurora_Australis_From_ISS.JPG*

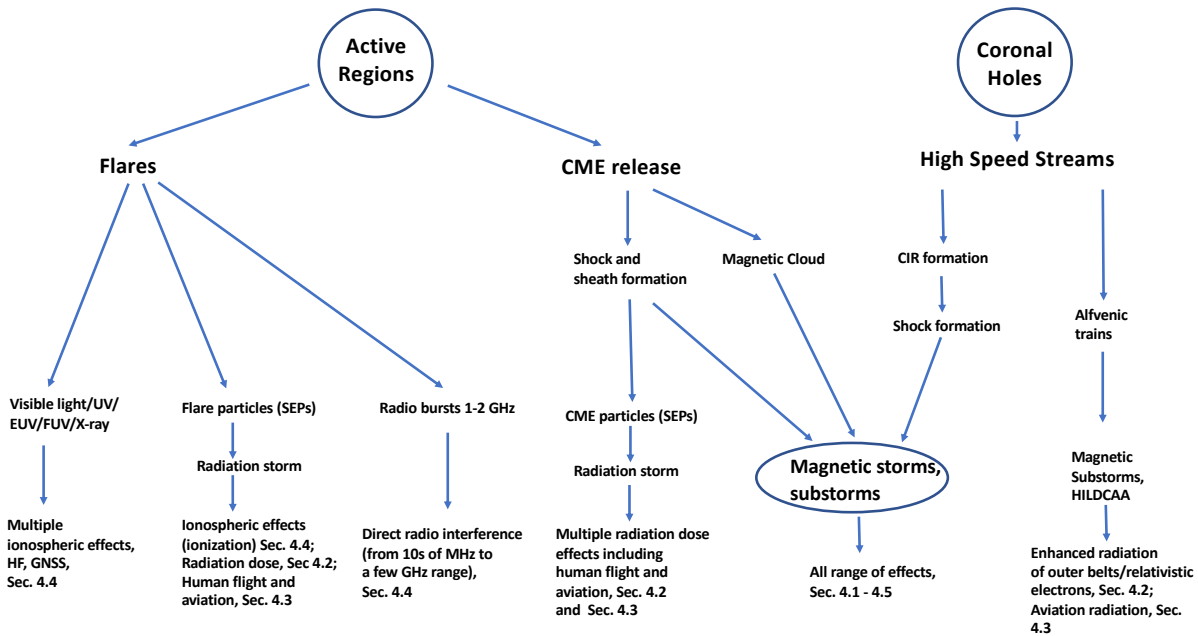

*Figure 2. Relationship between the main Space Weather processes (solar eruptive processes, and high-speed solar wind) and Space Weather effects/impacts (it is noted that CME/flare events can also be due to filament eruptions in quiet region and the boundaries of polar coronal holes, but most of the energetic eruptions originate from active regions). Different Space Weather effects/impacts could occur at the same time due to close association with eruptive processes, but could also occur independently, e.g. radiation storms and geomagnetic storms.*

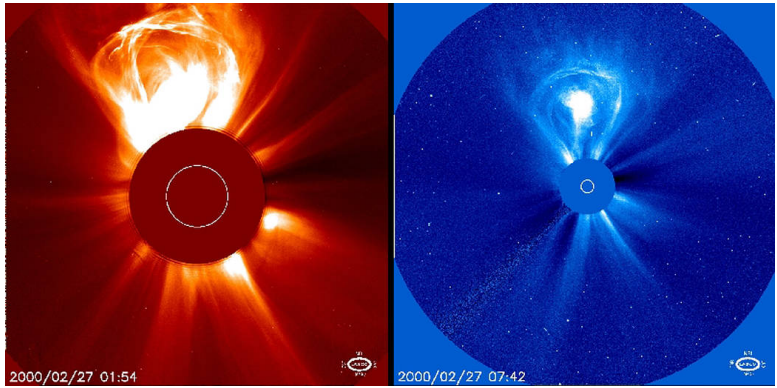

**Figure 3.** *An example of a Coronal Mass Ejection (CME). The CME grows larger as it moves away from the Sun and becomes an Interplanetary CME (ICME). Images are taken by NASA's SOHO on Feb 27, 2000. Credit: SOHO ESA & NASA. Source: https://www.nasa.gov/content/goddard/what-is-a-coronal-mass-ejection*

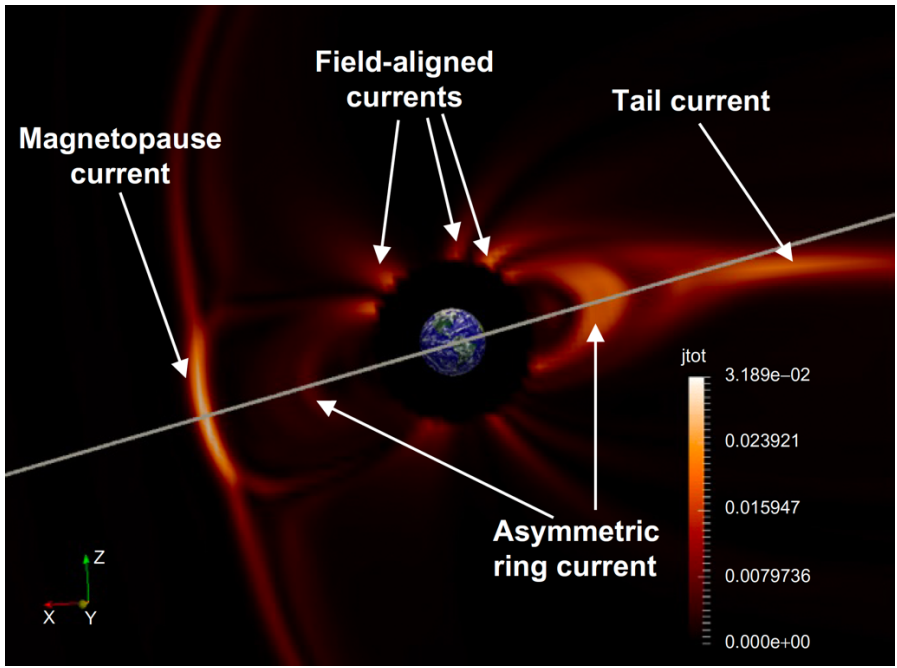

**Figure 4.** *A meridional cut through the global 3D simulation domain showing the results for the total current density for July 23, 2015 geomagnetic storm. The main magnetospheric current systems are presented: the asymmetric ring current, tail current, magnetopause current, and field-aligned currents (more visible in the regions of converging magnetic field). The sun is from the left. The gray line is the projection of a plane through the approximate geomagnetic equator in the inner magnetosphere. Current density is shown in μA/m². From Buzulukova et al., 2018b.*

**Space Weather: from solar origins to risks and hazards evolving in time**

**Footnotes**

1. Kristian Birkeland published a two volume book entitled "The Norwegian Aurora Polaris Expedition 1902-1903" where he introduced "Polar Elementary Storms" (Birkeland, 1908). Birkeland used his terrella experiments to show that "corpuscular rays" from the Sun could penetrate into the polar regions of the Earth. Analysis of Birkeland's contribution can be found in Egeland and Burke (2010). Akasofu's 1964 work was the first to introduce substorm morphology and provide a modern definition of substorm phases.

2. Although Edward E. Maunder is listed as the sole author of these two works, there is growing appreciation that his wife, astronomer and mathematician Annie S.D. Maunder, also contributed to the discovery of the statistical association between recurrent magnetic storms and solar activity. Please see for more details: https://mathshistory.st-andrews.ac.uk/Biographies/Maunder/

3. The list of near-Earth ICMEs for the period 1996-2022 could be found at https://izw1.caltech.edu/ACE/ASC/DATA/level3/icmetable2.htm

4. Curious readers may enjoy the Grand Archive of flare and CME cartoons as well as the list of relevant references at https://www.astro.gla.ac.uk/cartoons/index.html

5. The standard flare classification is based on a peak flux intensity (W/m2) in 1–8 A X-ray band. The three largest classes are X, M, C. The scale is logarithmic: a X1-class flare is 10 times more intense than a M1-class flare and a M1-class flare is 10 times more intense than a C1-class. Classification of X flares goes beyond X9 to X11, X28 (on November 4, 2003, the Sun's strongest X-ray flare on record), and to X100, X1000 and up for superflares from solar-like stars.

6. Only EUV radiation can ionize primary thermospheric species such as O, $O_2$ and $N_2$ (ionization energy thresholds are ~ 91, 103 and 80 nm respectively). An important thermospheric compound NO could be ionized by FUV Lyman-alpha radiation, contributing to the major ionization source at ~ 50-90 km, so-called D-layer of the ionosphere.

7. The current sources of International Sunspot Number and F10.7 are the Sunspot Index Data Centre at Royal Observatory of Belgium for ISN (https://www.sidc.be/silso/datafiles) and National Research Council Canada/Natural Resources Canada (https://www.spaceweather.gc.ca/forecast-prevision/solar-solaire/solarflux/sx-en.php).

8. In silicon, one electron-hole pair is produced for every 3.6 eV of energy lost by the ion. Therefore, a single 1 MeV ion could produce up to ~ 278,000 electron-hole pairs. See Bauman, 2005 for the details.

9. To reduce the risk of electrical discharges between different surface parts, a s/c needs to be designed and built so that all surface parts are electrically interconnected Although efforts have been made to make the s/c surfaces as conductive as possible, the spacecraft area still contains insulating materials, for example solar cell cover glass (Puthanveettil et al., 2014).

10. Frequencies ~ 10 GHz and up are likely not very sensitive to space weather. See also https://www.alsa.mil/News/Article/2532178/true-impacts-of-space-weather-on-a-ground-force/ for some examples.

11. The Fresnel zone defines the distance scale over which destructive interference can occur.

12. The effects of solar radio bursts on radars were first presented in secret reports during World War II documenting widespread disturbances of British radar systems operating at 55-85 MHz on February 26-28, 1942 (Hey, 1946).

13. Specification of the threshold for radio burst interference is receiver dependent.

14. https://www.unoosa.org/oosa/oosadoc/data/documents/2022/aac.105c.1l/aac.105c.1l.401_0.html See also Mann et al., 2018.

**Space Weather: from solar origins to risks and hazards evolving in time**

**Space Weather: from solar origins to risks and hazards evolving in time**

Space Weather: from solar origins to risks and hazards evolving in time

**Space Weather: from solar origins to risks and hazards evolving in time**

# Space Weather: from solar origins to risks and hazards evolving in time